\documentclass[twoside]{article}
\usepackage{graphicx}              

\def\real{{\rm I\kern-.2em R}}
\def\complex{\kern.1em{\raise.47ex\hbox{  $\scriptscriptstyle
|$}}\kern-.40em{\rm C}}

\title{An Adiabatic Theorem without a Gap Condition}
\author{J.~E.~Avron and A.~Elgart\\ Department of Physics, Technion, 32000 Haifa, Israel}

\begin{document}
\maketitle
\begin{abstract}
\noindent
The basic adiabatic theorems of classical and quantum mechanics are 
over-viewed and an adiabatic theorem in quantum mechanics without a gap condition
is described. 
\end{abstract}
\section{Classical Adiabatic Invariants}
 Consider a
(mathematical) pendulum whose period is slowly modulated, for example by
shortening the length of the pendulum,  fig 1,  \cite{arnold,lenard}.
\begin{center}
\includegraphics[height=4.cm]{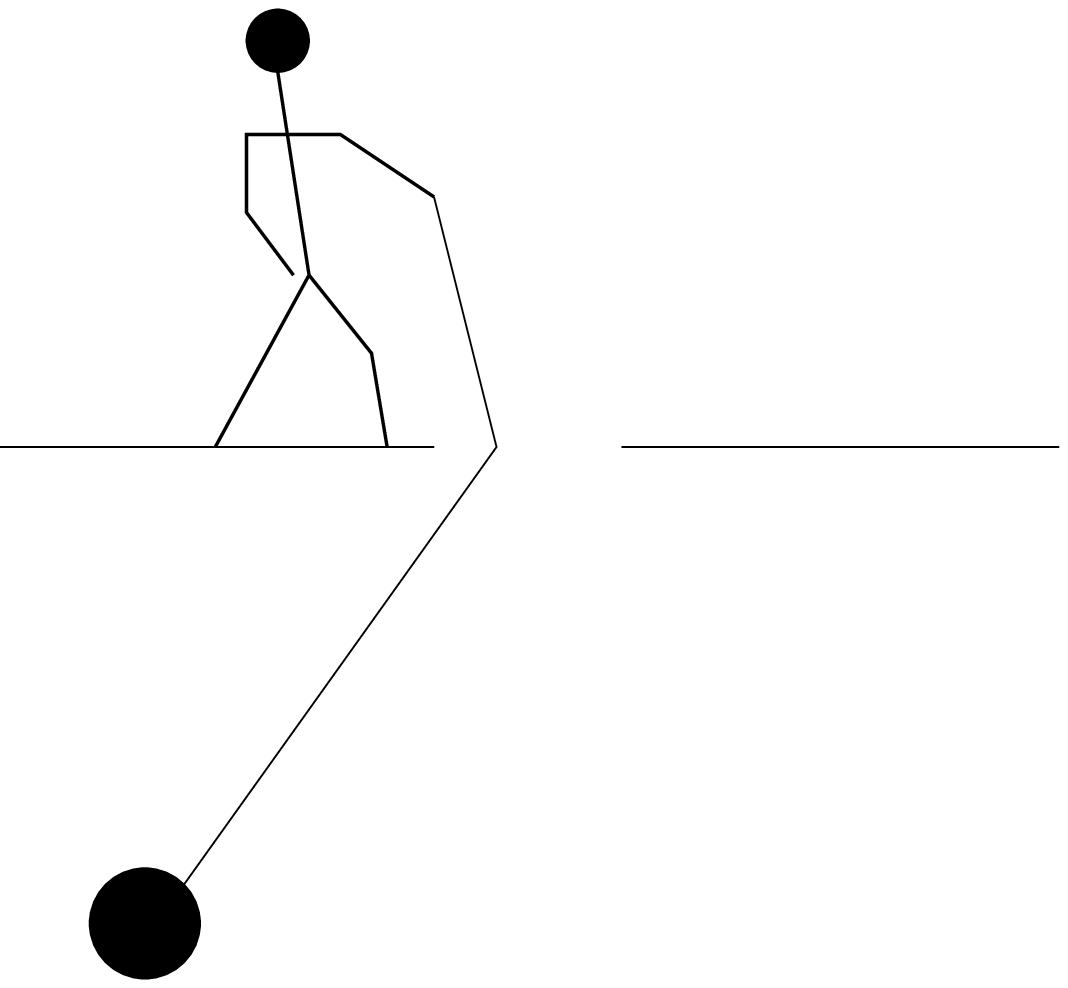}
\end{center}
\centerline{Figure 1:\hskip 0.125in An adiabatic pendulum}
 The Hamiltonian
describing the system is
\begin{equation}
H(s)= \frac{1}{2}\left( p^2 + \omega^2(s)\, x^2\right), \quad s=\frac{t}{\tau}.
\end{equation}
$t$ is the physical time, $\tau$ is the time scale. The adiabatic limit
is $\omega \tau>>1$. The period
$\omega(s)$ is  a smooth function which is time independent in the
past, $s<0$, and in the distant future, $s>1$. A graph showing a possible
variation of
$\omega(s)$ is shown in fig. 2.
\begin{center}
\includegraphics[height=4.cm]{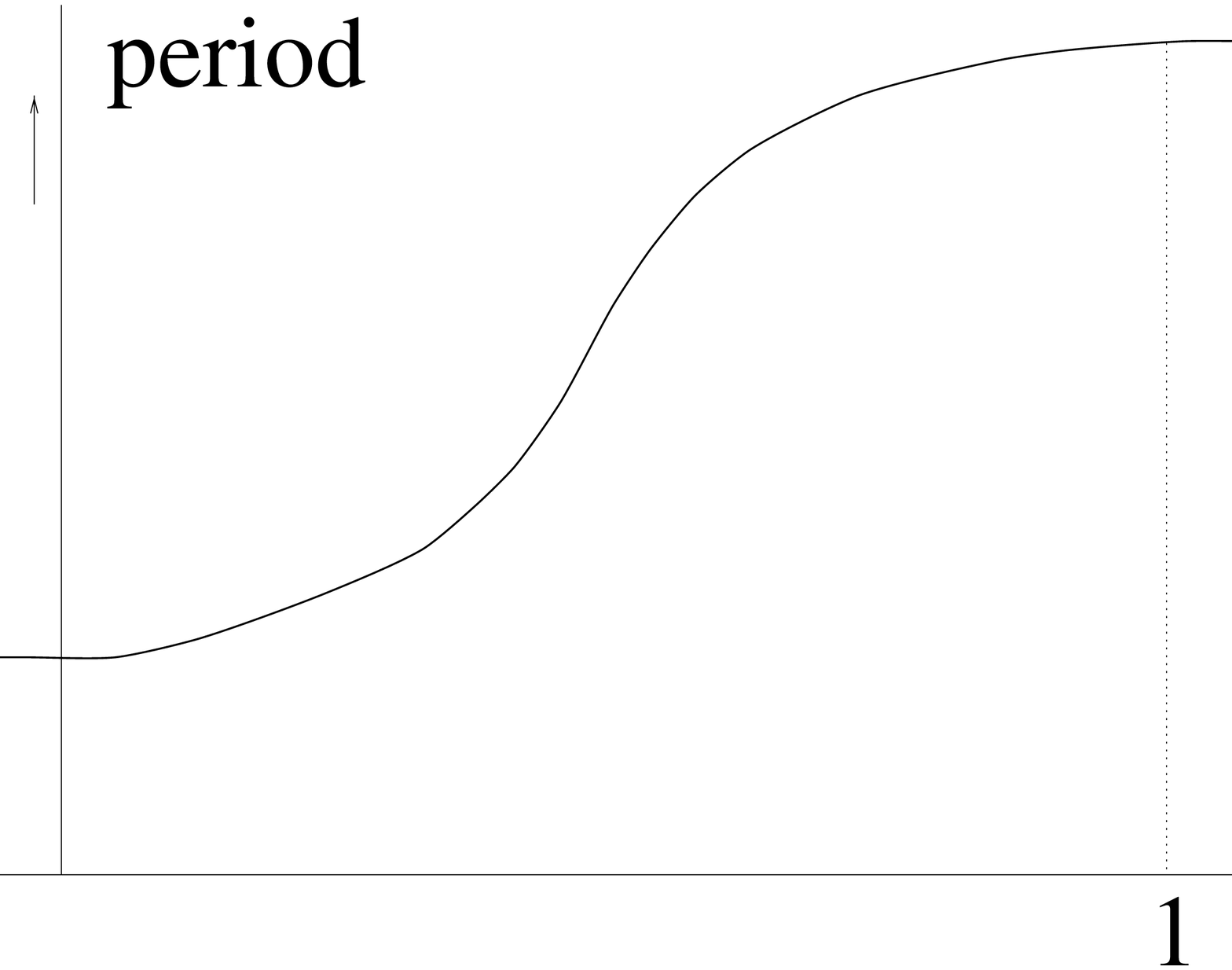}
\end{center}
\centerline{Figure 2:\hskip 0.125in An adiabatic variation}
An adiabatic invariant is an approximately conserved quantity  whose deviation
from constant can be made arbitrarily small for large $\tau$,
uniformly in $s$ and {\em for all times}. For the Harmonic oscillator the adiabatic invariant is
\begin{equation}
S(s)=\frac{ H(s)}{\omega(s)}.\label{s}
\end{equation}
The special properties of this particular combination of $H$ and
$\omega$ can be seen  from its equations of motion:
\begin{equation}
\dot S(s)=\frac{\dot \omega(s)}{2\,\omega^2(s)}\,\left(\omega^2(s)\, x^2
-p^2\right).
\end{equation}
$\dot S$ is compactly supported (because $\dot \omega$ is), and
appears to be $O(1)$ in $\tau$. But, for the (time independent) Harmonic oscillator the time average over one period
of the kinetic energy  equals the time average of the potential energy.
So, for large $\tau$, the change of $S$ in one period  is small:
$\langle\Delta S\rangle =O\left(\frac{1}{\tau}\right)$. Because of
this adiabatic invariants give precise information  on the long time
behavior even though the total variation in the Hamiltonian is finite.

A remarkable fact about adiabatic invariants is that {\em for large times}
the error is essentially exponentially small with $\tau$ if
$\omega(s)$ is smooth
\cite{lenard}:
\begin{equation}
|S(s)-S(0)|=O\left(\frac{1}{\tau^\infty}\right), \quad s>1.
\end{equation}
(The error is,  in general, not exponentially small for $0<s<1$.) In certain
circles an exponentially  small error is sometimes taken to be the
defining property of adiabatic invariant, so that  proving an adiabatic
theorem it taken to imply proving an exponentially small bound on the error.
This, to our opinion, is not a satisfactory definition of the notion of
adiabatic invariant, and instead we shall stick with the definition given above,
namely, that adiabatic invariants are conserved up to an error that is
uniformly bounded for all times, and can be made arbitrarily small with $\tau$.

A link of classical adiabatic invariants with quantum mechanics that was
emphasized by Ehrenfest \cite{ehrenfest} focused on the observation that
 adiabatic invariants are related to quantum numbers. For
the (time independent) Harmonic oscillator the particular combination of $H$ and
$\omega$ in Eq.~(\ref{s}) is a function of quantum numbers:
\begin{equation}
\frac{E}{\omega}= \hbar \left(n+\frac{1}{2}\right).
\end{equation}

\section{The Quantum Adiabatic Theorem}
Ehrenfest observation had much influence in the early days of quantum
mechanics,  and in particular motivated the work of Born and Fock
\cite{bf} on the adiabatic theorem of quantum mechanics.
In quantum theory one is interested is solving the initial value problem
\begin{equation}
i\,\partial_t \psi = H(s)\, \psi,\quad s=\frac{t}{\tau},\label{scrodinger}
\end{equation}
with $\psi$ a vector in Hilbert space and $H\left(\frac{t}{\tau}\right)$ a self
adjoint operator. We shall assume, as we did in the previous section, that
$H(s)$ is time independent in the past, $s<0$, and distant future, $s>1$, and is
a smooth operator valued function of $s$. In the case that $H(s)$ is an
unbounded operator, like the Schr\"odinger operator, the notion of smoothness
needs some clarification. We shall not get into this  here.

Changing variables one writes the initial value problem as
\begin{equation}
i\,\dot \psi =\tau\, H(s)\, \psi.
\end{equation}
The adiabatic limit is $\tau\to\infty$.
Adiabatic theorems in quantum mechanics relate the solutions of the initial
value problem to  spectral properties.

The oldest result of this kind is due to Born and Fock who studied
Hamiltonians  with {\em discrete and  simple} spectrum, fig. 3.
\begin{center}
\includegraphics[height=4.cm]{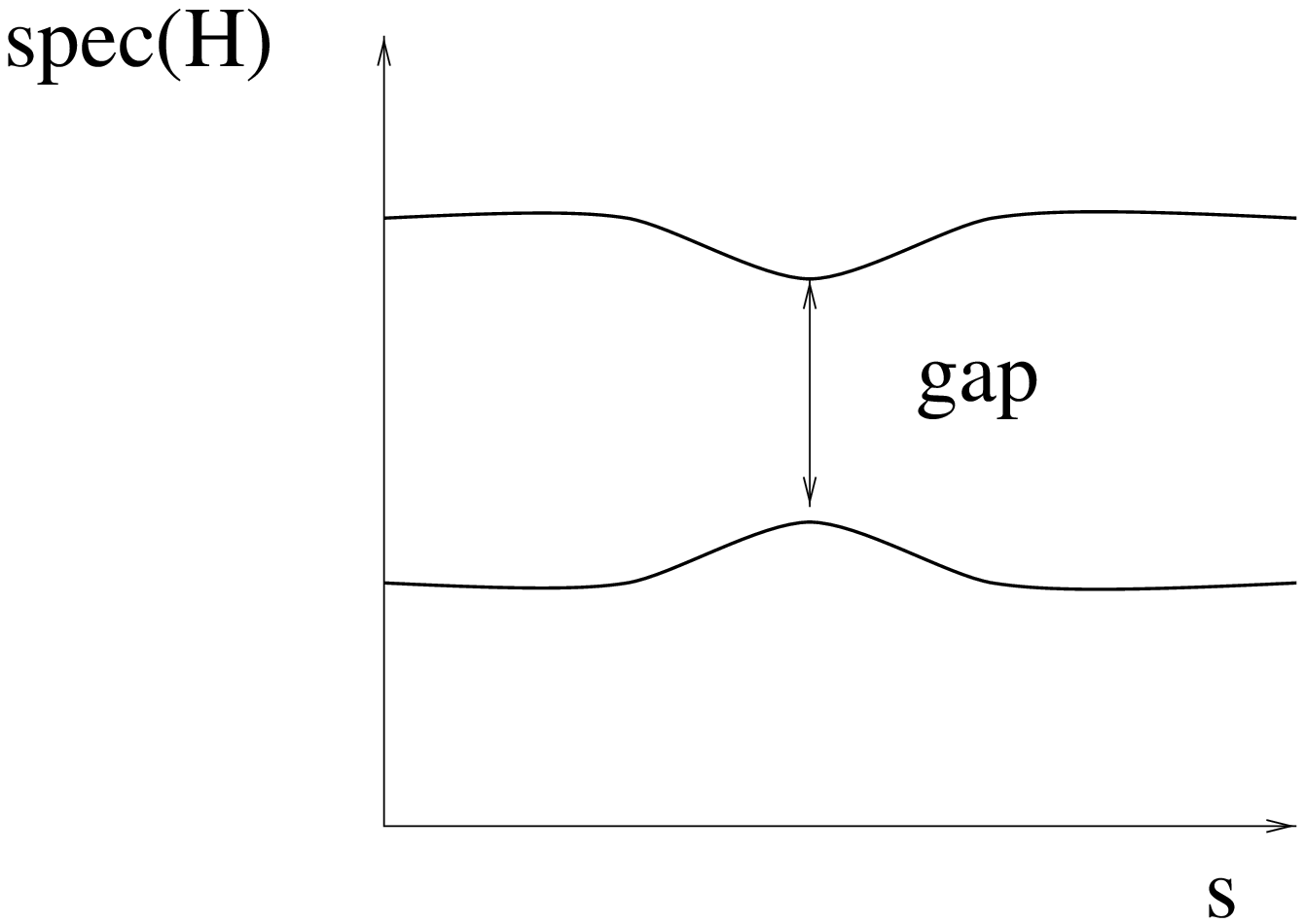}
\end{center}
\centerline{Figure 3:\hskip 0.125in Spectrum  in Born Fock Theory}

Born and Fock
showed that if the initial data are
$\psi(0)=\phi(0)$,  with  $\phi(0)$  an eigenvector of $H(0)$, then $\psi(s)$
is close to an eigenvector $\phi (s)$ of $H(s)$ with particular choice of phase:
\begin{equation}
\Vert \psi(s) -\phi(s)\Vert = O\left(\frac{1}{\tau}\right).\label{error}
\end{equation}
For large times, $s>1$, outside the support of $\dot H(s)$, much stronger
result hold: the error is essentially exponentially small in
$\tau$, see e.g. \cite{berry,joye,ks,nenciu}.

\section{The Adiabatic Theorem of Kato}

Kato generalized the result of Born and Fock. He showed that the assumption of
spectral simplicity of $H(s)$ can be removed, and so can the assumption that the
spectrum is discrete, fig. 4. These generalizations are important for
applications to atomic physics where some continuous spectrum is always
present, and degeneracies are ubiquitous. But, perhaps more importantly, Kato introduced an
essentially new method of proving the adiabatic theorem that we shall now
describe.
\begin{center}
\includegraphics[height=4.cm]{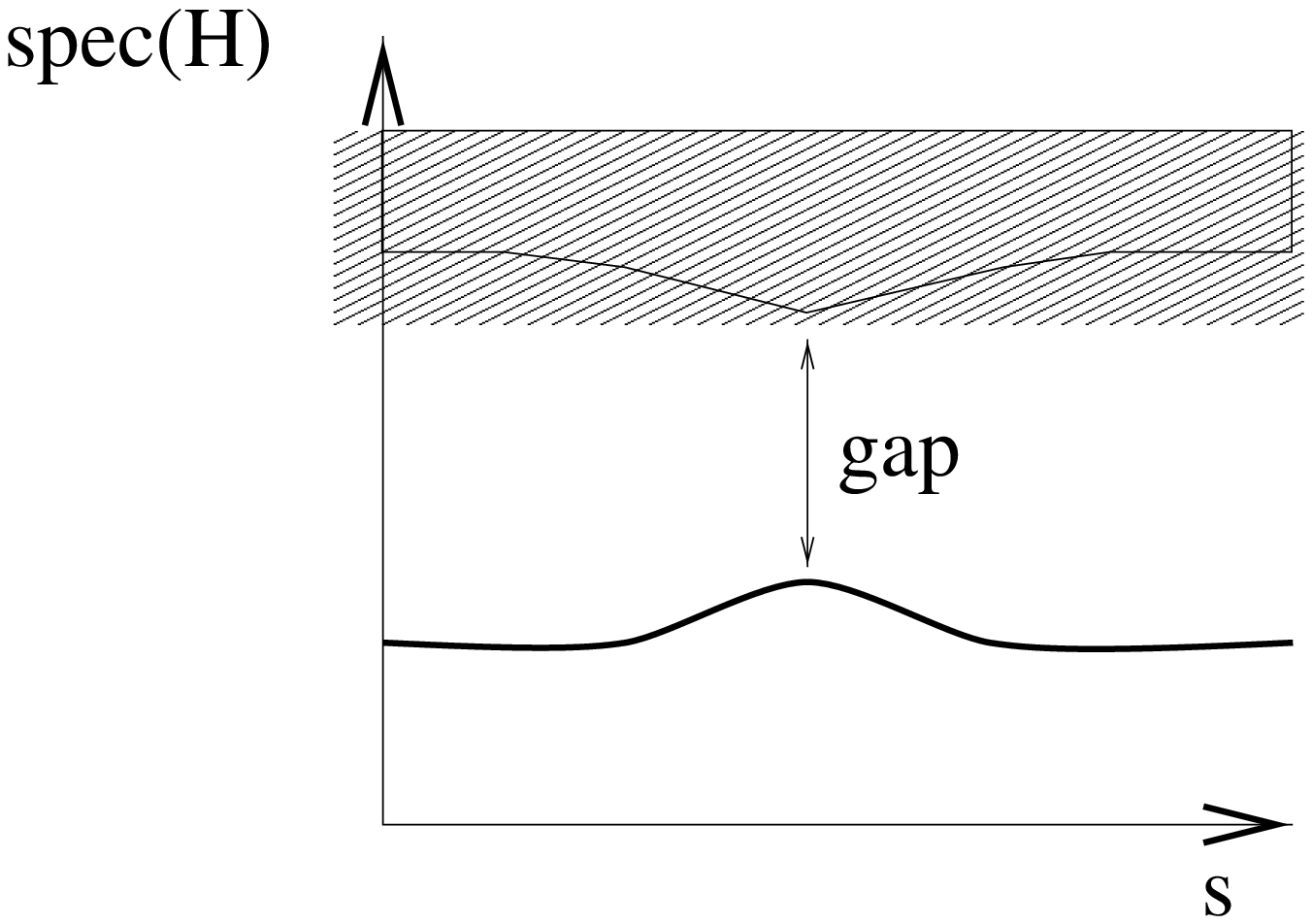}
\end{center}
\nopagebreak
\centerline{Figure 4:\hskip 0.125in Spectrum in Kato's Theory}
Kato's idea was to introduce a geometric evolution which satisfies the
adiabatic theorem without an error. That is,  a unitary
$U_a(s)$, so that:
\begin{equation}
U_a(s)\, P(0) = P(s)\, U_a(s),
\end{equation}
where $P(s)$ is a spectral projection for $H(s)$, and $U_a(0)=1$.
The second step is to compare the physical evolution, $U(s)$, generated by
\begin{equation}
i\,\dot U(s) =\tau\, H(s)\, U(s), \quad U(0)=1 .
\end{equation}
with $U_a$ and show that the two are close.

It turns out that both steps involve looking into commutator equations.
If we let $H_a(s)$ denote the generator of the geometric evolution $U_a(s)$,
it is not difficult to see that it must satisfy
\begin{equation}
\tau\, [H_a(s),P(s)]= i\,\dot P(s).
\end{equation}
Using the fact that for any projection $P$
\begin{equation}
P(s)\, \dot P(s)\,P(s)=0,
\end{equation}
one checks that
\begin{equation}
 H_a(s)=  H(s)+\frac{i}{\tau}\,[\dot P(s),P(s)],
\end{equation} solves the
commutator equation, with $H_a(s)$ which is manifestly close to $H(s)$.

To compare $U(s)$ and $U_a(s)$ let $\Omega(s)= U^*_a(s)\,U(s)$,
$\Omega(0)=1$.  Using the equation of motion one finds
\begin{equation}
\dot \Omega(s)= i\,\tau\,U^*_a(s)\left(H_a(s)-H(s) \right)\,U(s)=
-U^*_a(s)[\dot P(s), P(s)]\,U(s),
\end{equation}
which is compactly supported
(since $\dot P(s)$ is) and $O(1)$ in $\tau$. Now, like the situation
for the classical adiabatic invariants,  even though $\dot \Omega$ is
not small, the change in $\Omega$ is small.  This is where
 a second commutator equation enters. Suppose that the commutator
equation
\begin{equation}
[H(s),X(s)]=[\dot P(s), P(s)]\label{ce}
\end{equation}
has a smooth and bounded solution $X(s)$. Then,
\begin{eqnarray}
-\dot
\Omega(s)&=&U^*_a(s)[H(s),X(s)]\,U(s)=\nonumber \\
&=&U^*_a(s)\left(H_a(s)\,X(s)-X(s)\,H(s)\right)\,U(s)+
O\left(\frac{1}{\tau}\right)\nonumber
\\ &=&\frac{i}{\tau}\left( \dot U^*_a(s) \, X(s)\, U(s)+ U^*_a(s) \, X(s)\,
\dot U(s)\right) +O\left(\frac{1}{\tau}\right)\nonumber \\
 &=&\frac{i}{\tau} \left(\dot {\Big(U^*_a(s) \, X(s)\, U(s)\Big)}- U^*_a(s)
\, \dot X(s)\, U(s)\right) +O\left(\frac{1}{\tau}\right).
\end{eqnarray}
From this it follows that $\Omega(s)-1=O\left(\frac{1}{\tau}\right)$.

The gap condition is a condition for the solvability of the commutator
equation, Eq.~(\ref{ce}). Indeed, suppose there is a gap in the spectrum so that
the spectral projection
$P$ is associated with a contour $\Gamma$ in the complex plane that lies
entirely in the resolvent set, Fig. 5.
\begin{center}
\includegraphics[height=4.cm]{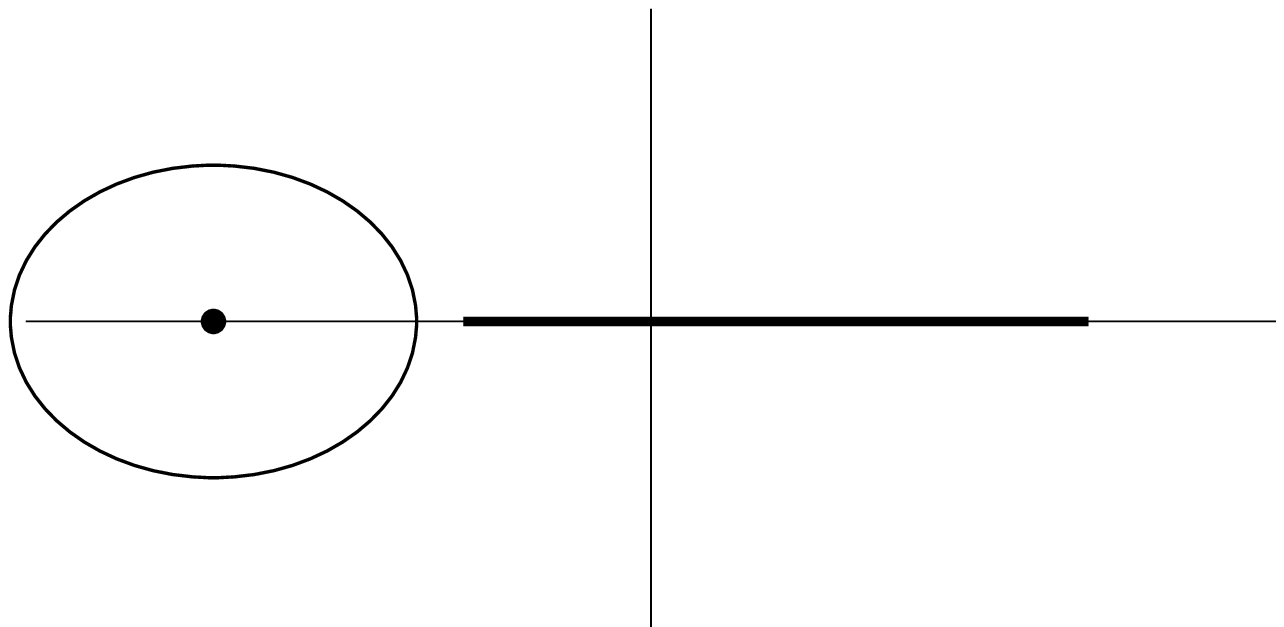}
\end{center}\bigskip
\nopagebreak
\centerline{Figure 5:\hskip 0.125in A contour $\Gamma$ in the Complex Plane}
A solution to the
commutator equation is
\begin{equation}
X(s)=\frac{1}{2\pi i}\, \int_\Gamma\, R(s,z) \, \dot P(s)\, R(s,z)\,
dz.\label{x}
\end{equation}
And, as usual $R(s,z)= (H(s)-z)^{-1}$ is  bounded for $z\in\Gamma$.
If the size of the gap is $g$ then
\begin{equation}
\Vert X(s)\Vert =O\left(\frac{1}{g }\right).\label{b}
\end{equation}
Using Kato's method various adiabatic theorems have been proven see e.g.
\cite{ahs,asy,ks,nenciu}.

\section{The Role of the Gap Condition}
The adiabatic theorem described in the previous section  relied on a gap
condition. How serious is this?

In the proof of Kato the gap condition guarantees the existence of a
bounded solution to the commutator equation given by
$X(s)$ of Eq.~(\ref{x}).
The bound Eq.(\ref{b})  blows up at the gap shrinks to zero and there is no {\it {a-priori}}
bounded solution to the commutator equation. This suggests that the
gap condition is essential.

A second argument leading to the same conclusion is a dimensional argument. The
adiabatic limit  needs a intrinsic time scale so that $\tau$ can be
measured in dimensionless units. Otherwise the notion of large $\tau$
 depends on a choice of a unit and is meaningless. In the case
of the classical Harmonic oscillator the intrinsic time scale is set by
$\omega$. In the quantum case, a  gap and Planck constant dictates an intrinsic
time scale. In the absence of a gap, this time scale is lost. This
suggests that the gap condition is essential and there should be no {\em
general} adiabatic theorem in its absence.

Let us now describe two arguments that say the opposite. The first refers
once again to the work of Born and Fock. Born and Fock (and also Kato) considered
the more delicate adiabatic theorem for crossing energy levels, Fig. 6, and 
proved an adiabatic theorem in this case 
\begin{center}
\includegraphics[height=4.cm]{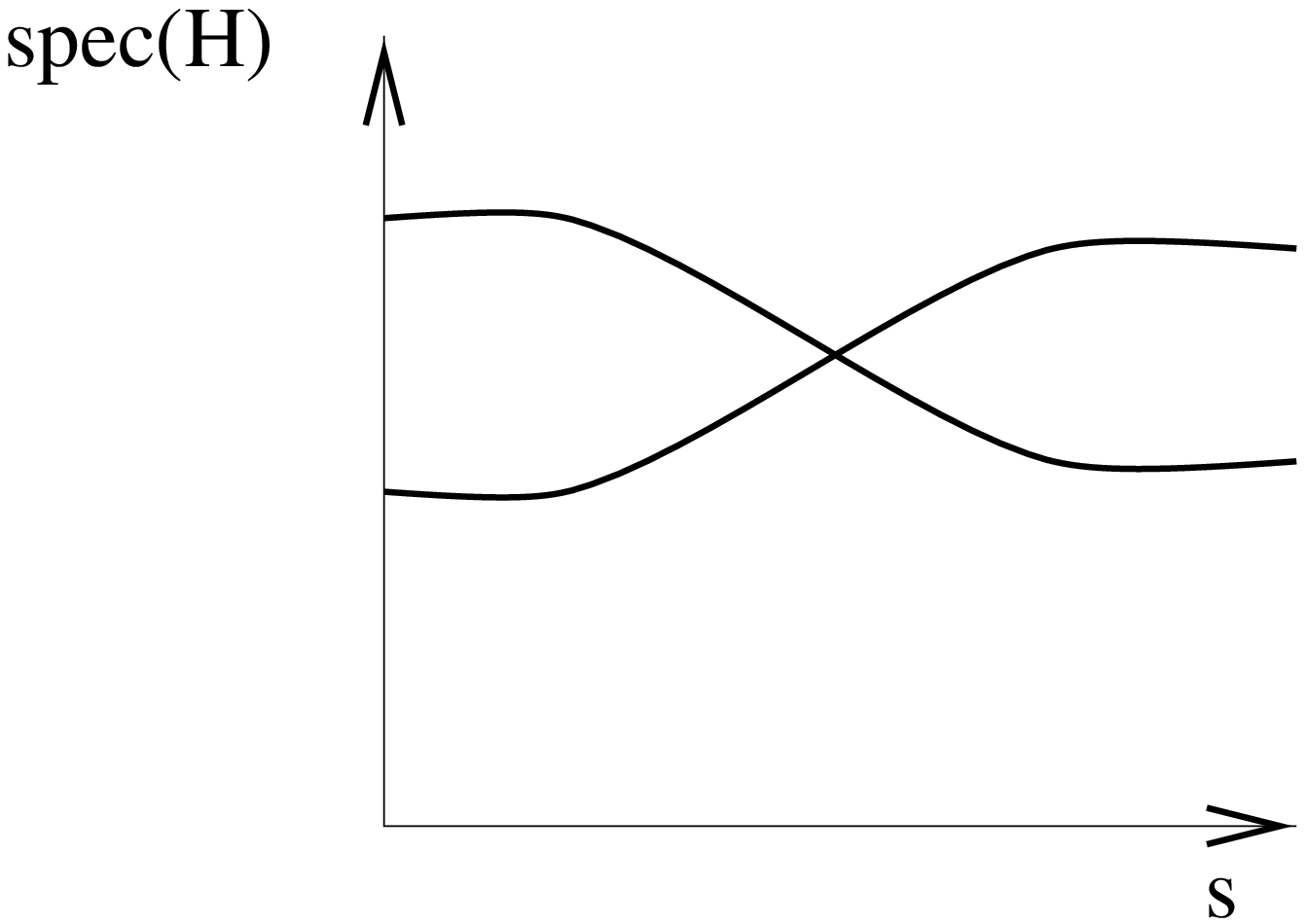}
\end{center}
\bigskip
\centerline{Figure 6:\hskip 0.125in Crossing Eigenvalues in Born Fock Theory}
Since the energy levels cross, the gap closes. For the
case of linear crossing Born and Fock showed that  Eq.~(\ref{error})
is replaced by 
\begin{equation}
\Vert \psi(s) -\phi(s)\Vert = O\left(\frac{1}{\sqrt\tau}\right).
\end{equation}
The  time scale for this problem is dictated by the slope of the energy curves
at the crossing point. 
These results have since been considerably strengthened and 
extended \cite{hagedorn}. 
This suggests that a gap conditions controls the rate at which the adiabatic
limit is approached, but an adiabatic theorem does not really require a gap
condition.

A second argument supporting the view that a gap condition is only technical is
a physical argument. Gaps in the spectrum are indeed prevalent in quantum
mechanical systems, but they are no gaps in quantum electrodynamics: The
interaction with radiation eliminates the gaps.  Suppose that a charged quantum mechanical system, initially
at the ground state, is slowly rotated. The adiabatic theorem would fail if the
 number of photons generated by the slow rotation does not
go to zero in the adiabatic limit. Let us
estimate this number
\footnote{We owe this argument to A. Ori.}. The power radiated by a charged system
in classical electrodynamics is proportional to  the acceleration squared, i.e.
to
$\tau^{-4}$. Hence the  total radiated energy is of the order
$\tau^{-3}$. Since  a typical radiated photon will, presumably,
have frequency of order
$\frac{1}{\tau}$ the number of radiated photons is of order $\tau^{-2}$.
This goes to zero in the adiabatic limit. This argument, in spite of 
its shortcomings, suggests that the gap
condition, at least in the context of QED, is not really essential.

\section{Removing the Gap Condition}
A general adiabatic theorem without a
gap condition was given in
\cite{ae}. The point is that all the adiabatic theorem really
needs is a distinguished  smooth family of finite dimensional spectral projections,
so that the adiabatic evolution has a distinguished subspace to follow.
The proof works for eigenvalues embedded in some essential spectrum, or
for eigenvalues at the threshold of essential spectrum,  as one would 
expect to find in QED, fig. 7.
It is essential
for this result that the distinguished spectral subspace is {\em finite
dimensional}.  Let us begin by stating the
theorem:\medskip\hfil\break\noindent {\em Theorem: Suppose that
$P(s)$ is a finite rank spectral projection, which is at  least twice
differentiable (as a bounded operator), for the self-adjoint Hamiltonian
$H(s)$, which is bounded and differentiable  for all $s\in[0,1]$. Then, the
evolution of the initial state
$\psi(0)\in Range P(0)$, according to Eq.~(\ref{scrodinger}), is such that in
the adiabatic limit
$\psi (s)\in Range P(s)$ for all $s$.
}
\medskip
\begin{center}
\includegraphics[height=4.cm]{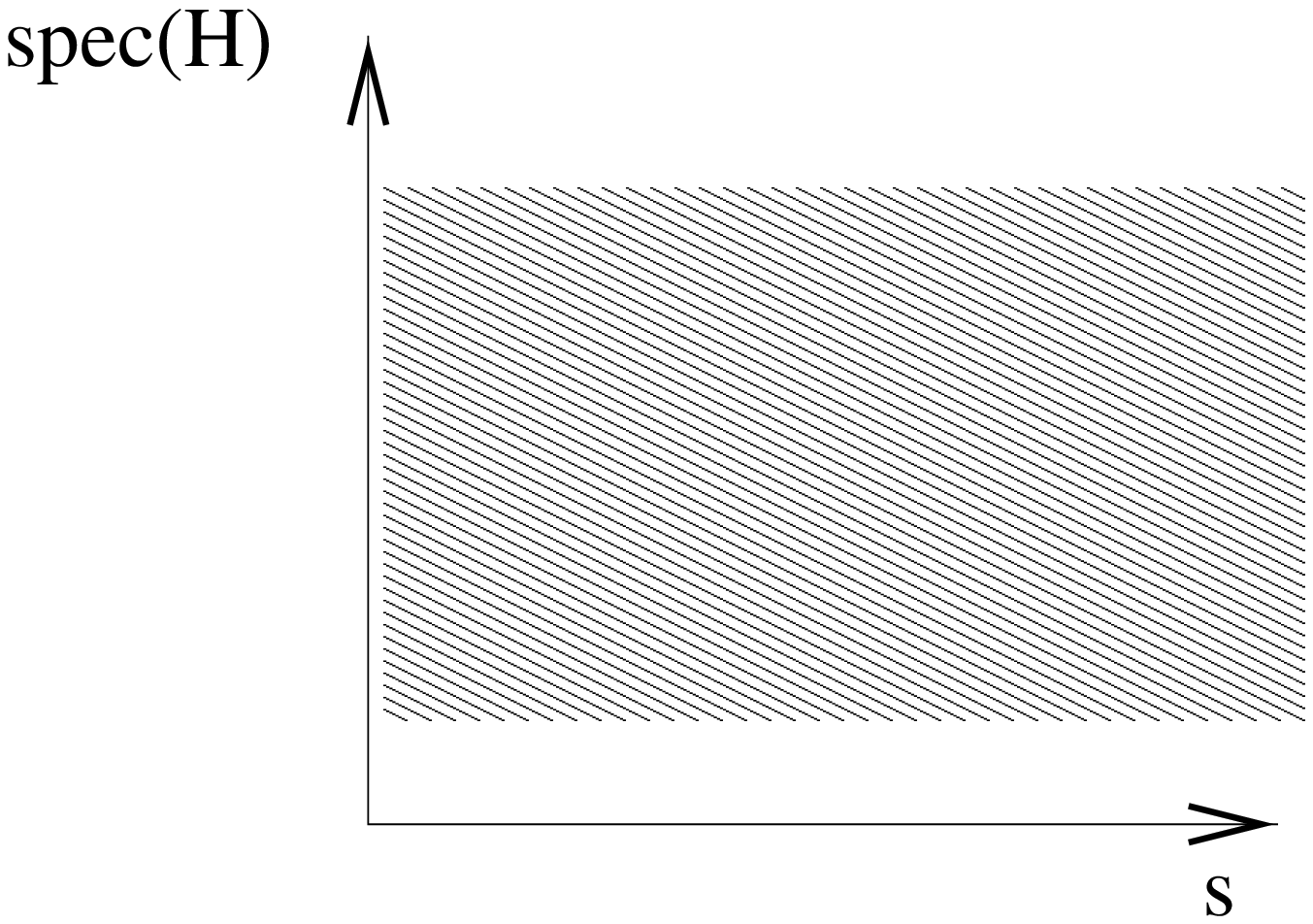}
\end{center}
\nopagebreak
\centerline{Figure 7:\hskip 0.125in An Eigenvalue at Threshold}
Remarks: The theorem is stated for bounded self adjoint operators $H(s)$. As it
stands it does not apply to Schr\"odinger operator.
The extension to unbounded operators is a technical problem which can be
handled by known functional analytic methods. We choose not to phrase the result
for the general case for several reasons. The first is that the technical
issues will obscure the basic idea which is simple. The second is that
the essence of the adiabatic theorem is an infrared problem. 
The unboundedness of Schr\"odinger operators is an ultraviolet
problem. It is a conceptual advantage to keep the two issues separate.

The basic idea  is to replace Kato's commutator equation,
Eq.~(\ref{ce}), by a definition of a new quantity $Y(s)$:
\begin{equation}
[H(s),X(s)]=[\dot P(s), P(s)]+Y(s),
\end{equation}
and take $X(s)$ to be
\begin{equation}
X_\Delta(s)=\frac{1}{2\pi i}\,\int_\Gamma  dz\, (1-F_\Delta(s))\,R(z,s)\,
\dot P(s)
R(z,s)\,(1-F_\Delta(s)).
\end{equation}
where $F_\Delta(s)$ is an approximate characteristic function  of $H(s)$, which
is $\Delta$ localized near the relevant eigenvalue, whose range is in $Range
\,P_\perp(s)$.
$X_\Delta (s)$ is bounded, by construction, for $\Delta>0$ and its norm
diverges as $\Delta\to 0$. At the same time, and this is the crucial point,
$\Vert Y_\Delta(s)\Vert\to 0$ provided $P(s)$ is finite dimensional. Chasing the argument of Kato one
then finds that the adiabatic theorem holds, and the price one has to pay for
the absence of a gap is the loss of control on the rate  at which the adiabatic
limit is approached. Instead of Eq.~(\ref{error}) on gets
\begin{equation}
\Vert \psi(s) -\phi(s)\Vert = o\left(1\right).
\end{equation}
That is, the error can be made arbitrarily small with $\tau$, but the rate is
undetermined.

We conclude with an interpretation of the result. For an isolated eigenvalue the gap in the spectrum
protects against tunneling out of the spectral subspace. In the case that the
eigenvalue in question is embedded in essential spectrum there is no gap to
protect against tunneling out. But, since the essential spectrum is
associated with eigenfunctions supported near infinity, there is small overlap
with the eigenfunction in question, and the protection against tunneling comes
from this fact.
\section{What Has Been Left Out}
Adiabatic theorems of classical and quantum mechanics are a developed 
subject with rich and fertile history. In this short overview, based 
an a talk by one of us, we  reviewed  a small corner of this field, 
the one close to its foundations and characterized by elementary results.
There are many beautiful and sophisticated results that 
we did not have the opportunity to review. These include: Classical 
adiabatic invariants for integrable systems to all orders \cite{arnold,lenard,lm}; 
Adiabatic invariants for chaotic 
systems \cite{ott,br,jarzinski}; Quantum adiabatic theorems to all orders
\cite{berry,ks,nenciu,joye}; Landau-Zener 
formulas \cite{js,joye}; Adiabatic invariants in scattering 
theory\cite{nt}; Adiabatic 
invariants in $C^{* }$ algebras and models of quantum fields \cite{ds} 
and geometry and adiabatic curvature 
\cite{berry2,k}.

\section*{Acknowledgments}  This work
was partially supported by a grant from the Israel Academy of Sciences, 
the Deutsche
Forschungsgemeinschaft, and by the Fund for Promotion of Research at the Technion.


\end{document}